\documentclass[12pt]{article}

\usepackage[letterpaper,hmargin=1in,vmargin=1in]{geometry}

\usepackage{graphicx,epstopdf,amsmath,amsfonts,amssymb,cite}
\usepackage[hypertex]{hyperref}
\usepackage[small,compact]{titlesec} 

\def\be{\begin{equation}}
\def\ee{\end{equation}}
\def\ba{\begin{eqnarray}}
\def\ea{\end{eqnarray}}
\def\ge{\mathrel{\raise.3ex\hbox{$>$\kern-.75em\lower1ex\hbox{$\sim$}}}}
\def\la{\mathrel{\raise.3ex\hbox{$<$\kern-.75em\lower1ex\hbox{$\sim$}}}}

\def\blfootnote{\xdef\@thefnmark{}\@footnotetext}

\def\simgt{\mathrel{\raise.3ex\hbox{$>$\kern-.75em\lower1ex\hbox{$\sim$}}}}
\def\simlt{\mathrel{\raise.3ex\hbox{$<$\kern-.75em\lower1ex\hbox{$\sim$}}}}

\newcommand{\fr}[2]{\frac{#1}{#2}}

\newcommand{\nc}{\newcommand}

\nc{\gone}{\bar g_{\pi NN}^{(1)}}
\nc{\gzero}{\bar g_{\pi NN}^{(0)}}
\nc{\al}{\alpha}
\nc{\ga}{\gamma}
\nc{\de}{\delta}
\nc{\ep}{\epsilon}
\nc{\ze}{\zeta}
\nc{\et}{\eta}
\nc{\ka}{\kappa}
\nc{\rh}{\rho}
\nc{\si}{\sigma}
\nc{\ta}{\tau}
\nc{\up}{\upsilon}
\nc{\ph}{\phi}
\nc{\ch}{\chi}
\nc{\ps}{\psi}
\nc{\om}{\omega}
\nc{\Ga}{\Gamma}
\nc{\De}{\Delta}
\nc{\La}{\Lambda}
\nc{\Si}{\Sigma}
\nc{\Up}{\Upsilon}
\nc{\Ph}{\Phi}
\nc{\Ps}{\Psi}
\nc{\Om}{\Omega}
\nc{\ptl}{\partial}
\nc{\del}{\nabla}
\nc{\ov}{\overline}
\nc{\newcaption}[1]{\centerline{\parbox{15cm}{\caption{#1}}}}
\nc{\us}{U(1)$_S$}

\def\beq{\begin{equation}}
\def\eeq{\end{equation}}
\def\bmat{\begin{displaymath}}
\def\emat{\end{displaymath}}
\def\bear{\begin{eqnarray}}
\def\eear{\end{eqnarray}}
\def\ba{\begin{eqnarray}}
\def\ea{\end{eqnarray}}
\def\bery{\begin{array}}
\def\ery{\end{array}}
\def\bit{\begin{itemize}}
\def\eit{\end{itemize}}
\def\ben{\begin{enumerate}}
\def\een{\end{enumerate}}
\def\btab{\begin{tabular}}
\def\etab{\end{tabular}}
\def\btbl{\begin{table}}
\def\etbl{\end{table}}
\def\bfig{\begin{figure}[htb]}
\def\efig{\end{figure}}
\def\bpic{\begin{picture}}
\def\epic{\end{picture}}

\def\ga{\mathrel{\raise.3ex\hbox{$>$\kern-.75em\lower1ex\hbox{$\sim$}}}}
\def\la{\mathrel{\raise.3ex\hbox{$<$\kern-.75em\lower1ex\hbox{$\sim$}}}}
\def\gappeq{\mathrel{\rlap {\raise.5ex\hbox{$>$}}
{\lower.5ex\hbox{$\sim$}}}}
\def\lappeq{\mathrel{\rlap{\raise.5ex\hbox{$<$}}
{\lower.5ex\hbox{$\sim$}}}}

\def\gyr{{\rm \, G\kern-0.125em yr}}
\def\mev{{\rm \, Me\kern-0.125em V}}
\def\gev{{\rm \, Ge\kern-0.125em V}}
\def\tev{{\rm \, Te\kern-0.125em V}}

\newcommand{\ud}{\textrm{d}}

\newcommand{\vev}[1]{\left\langle #1\right\rangle}
\newcommand{\twopt}[1]{\vev{#1\,,\dots}}

\renewcommand{\bar}{\overline}
\newcommand{\tr}{\textrm{tr}}

\newcommand{\dslash}[1]{\displaystyle{\not}{#1}}

\newcommand{\MP}{M_{\textrm{pl}}}

\newcommand{\rst}[1]{\raise+.6ex\hbox{#1}}
\newcommand{\T}{\textrm{\rst{{\tiny T}}}}
\newcommand{\TT}{\textrm{\rst{\tiny TT}}}

\newcommand{\LambdaHL}{\Lambda_{\textrm{HL}}}
\newcommand{\LambdaUV}{\Lambda_{\textrm{UV}}}

\begin{document}

\begin{titlepage}

\setcounter{page}{1}

\vspace*{0.2in}

\begin{center}

\hspace*{-0.6cm}\parbox{17.5cm}{\Large \bf \begin{center}

On Lorentz violation in Ho\v rava-Lifshitz type theories

\end{center}}

\vspace*{0.5cm}
\normalsize

\vspace*{0.5cm}
\normalsize

{\bf Maxim Pospelov$^{\,(a,b)\ast}$ and Yanwen Shang$^{\,(a)\dag}$ }

\smallskip
\medskip

$^{\,(a)}${\it Perimeter Institute for Theoretical Physics, Waterloo,
ON, N2J 2W9, Canada}

$^{\,(b)}${\it Department of Physics and Astronomy, University of Victoria, \\
     Victoria, BC, V8P 1A1 Canada}

\smallskip
\end{center}
\vskip0.2in
%\centerline{\large\bf Abstract}
\begin{abstract}
We show that coupling the Standard Model to a Lorentz
symmetry violating sector may co-exist with viable 
phenomenology, provided that the interaction between the two 
is mediated by higher-dimensional operators. In particular,
if the new sector acquires anisotropic scaling behavior above 
a ``Ho\v rava-Lifshitz'' energy scale $\LambdaHL$ and couples to the 
Standard Model through interactions suppressed by $\MP$, 
the transmission of the Lorentz violation
into the Standard Model is protected by the ratio 
$\LambdaHL^2/\MP^2$. A wide scale separation, $\LambdaHL\ll \MP$, 
can then make Lorentz-violating terms 
in the Standard Model sector within experimental bounds 
without fine-tuning. 
We first illustrate our point with a toy example of 
Lifshitz-type neutral fermion 
coupled to photon via the magnetic moment operator, and then 
implement similar proposal for the Ho\v rava-Lifshitz 
gravity coupled to conventional Lorentz-symmetric matter fields.
We find that most radiatively induced Lorentz violation
can be controlled by a large scale separation, 
but the existence of instantaneously 
propagating non-Lifshitz modes in gravity can  cause a certain 
class of diagrams to remain quadratically divergent above $\LambdaHL$.
Such problematic quadratic divergence, however, can be removed by extending 
the action with terms of higher Lifshitz dimension,
resulting in a completely consistent setup that can cope with
the stringent tests of Lorentz invariance.
\end{abstract}

\vskip0.7in
~\\November~14,~2011
\vskip0.1in
~\\\rule{.5\linewidth}{0.1mm}
{\\\small $^\ast$pospelov@uvic.ca \\ $^\dag$yshang@perimeterinstitute.ca}

\end{titlepage}

%%%%%%%%%%%%%%%%%%%%%%%%%%%%%%%%%%%%%%%%%%%%%%%%%%%%%%%%%%%%%%%%%%%%%%%%%%%%%%%%
%%%%%%%%%%%%%%%%%%%%%%%%%%%%%%%%%%%%%%%%%%%%%%%%%%%%%%%%%%%%%%%%%%%%%%%%%%%%%%%%
%                           I N T R O D U C T I O N
%%%%%%%%%%%%%%%%%%%%%%%%%%%%%%%%%%%%%%%%%%%%%%%%%%%%%%%%%%%%%%%%%%%%%%%%%%%%%%%%
%%%%%%%%%%%%%%%%%%%%%%%%%%%%%%%%%%%%%%%%%%%%%%%%%%%%%%%%%%%%%%%%%%%%%%%%%%%%%%%%
\section{Introduction}
\label{sec:introduction}

Lorentz symmetry, and its universality with respect to propagation and interaction of different types of 
particles, is a very well-established symmetry of nature. Stringent constraints are derived on 
the parameters of effective Lagrangian that encode possible departures from Lorentz symmetry 
\cite{Colladay:1996iz,Coleman:1998ti}. 
Existing models of Lorentz symmetry breaking did not go far beyond the 
effective Lagrangian description, and the idea that either a vector or the gradient of a scalar 
field condense at intermediate or low energy while restoring the Lorentz symmetry at high energies 
\cite{Kostelecky:1989jw,Eling:2004dk,ArkaniHamed:2003uy}
so far has not found any reasonable ultraviolet (UV) completion. Even more, it is  not fully understood whether 
such completions exist in principle. 

It is also conceivable that Lorentz symmetry is somehow broken by the UV physics, and for example quantum 
gravity is often being tauted as being capable of causing that (see {\em e.g.} \cite{AmelinoCamelia:1997gz}). 
If Lorentz violation (LV) is indeed a 
UV-related phenomenon, then there is a significant conceptual hierarchy problem. 
One would expect that LV should manifest itself in the lowest dimensional operators. 
Since the set of such operators starts from dimensions
3 and 4 \cite{Colladay:1996iz,Coleman:1998ti}, one should naively expect that 
the strength of LV interactions is of the order of 
$\Lambda_{ \rm LV}$ for dimension 3 operators,  and $O(1)$ for dimension 4. 
Several mechanisms of protecting higher-dimensional LV operators from 
``leaking'' into the lower dimensional ones have been proposed and 
partially summarized in 
\cite{Bolokhov:2007yc}. 

The localization of LV to higher-dimensional operators can occur in 
various ways.
For example, Ref. \cite{Myers:2003fd} assumed that operators 
responsible for Lorentz violation 
are tensors of a higher rank and irreducible, 
and therefore their appearance in dimension 3 and 4 operators
is prohibited.  Refs. \cite{GrootNibbelink:2004za,Bolokhov:2005cj} 
argue that supersymmetrization of the Standard Model (SM) leads to 
automatic elimination 
of lower dimensional LV operators. The soft-breaking terms allow this leakage into lower dimensions to happen, 
but in a controllable way: {\em e.g.} the coefficients of dimension 4 operators are induced 
by the dimension 6 operators:
\be
c^{(4)}_{ \rm LV} \sim m_{\rm soft}^2 c^{(6)}_{\rm LV} 
\sim \fr{m_{\rm soft}^2}{\Lambda^2_{\rm LV} }.
\label{susy}
\ee
If there is a wide enough scale separation between the SUSY breaking mass and the 
high-energy scale where LV originates, $m_{\rm soft}\ll \Lambda_{\rm LV}$, the existence of Lorentz breaking 
can be made consistent with the variety of experimental constraints. Dimension 4 coefficients  $c^{(4)}_{\rm LV}$
induce a difference between propagation speed 
for different particles, limited by the most stringent constraints 
to be at the level of $10^{-23}$ (see {\em e.g.} \cite{Gagnon:2004xh}),
which is perfectly safe, for example,  if $m_{\rm soft}$ is at the weak scale and $\Lambda_{\rm LV}$ is close to Planck scale.

In this paper we examine another generic but very different way of 
protecting against  LV leaking into the SM sector. 
Consider a LV-sector that couples to the SM via a
power-suppressed interaction: 
\be
\label{powerlike}
 \fr{1}{M^{n+k-4}}  O^{(n)}_{\rm LV}O^{(k)}_{\rm SM},
\ee
where $O^n_{LV}$ and  $O^k_{SM}$ are some operators from LV and SM 
sectors of dimensions 
$n$ and $k$ respectively and $n+k \geq 5$, and $M$ is a very high
energy scale. Being power-suppressed, this operator 
would typically generate a power-divergent loop integral. For example, when 
$n=1$ and $k=4$, integrating out fields in the LV sector is likely to
generate a quadratic divergence leading to an LV term in the SM 
as:
\be
\label{radcorr}
 \fr{1}{M}  O^{(1)}_{\rm LV}O^{(4)}_{\rm SM} \to \fr{\LambdaUV^2}{M^2}O^{(4)}_{\rm SM,LV}.
\ee
Theories of this kind are usually not considered viable 
on the phenomenological ground. The induced LV term is 
generically of order 1 since naturally
$\LambdaUV\sim M$.  However, particularly interesting cases exist
when the loops in the LV sector
are stabilized at high energy through certain mechanism
so that $\LambdaUV$ gets replaced by a well-defined 
physical scale that can be separated far from $M$. In the latter
case, the induced LV terms as in \eqref{radcorr} can be made
arbitrarily small.

A well-known class of mechanisms of such kind is introducing 
higher-derivative terms in the interactions 
or propagators, which improves the convergence of loop 
integrals. Examples include the 
non-commutative field theories \cite{Filk:1996dm,Minwalla:1999px}, the so-called Lee-Wick theories 
\cite{Lee:1969fy,Grinstein:2007mp} and Ho\v rava-Lifshitz type theories \cite{Lifshitz,Horava:2009uw}. 
In the last example, the following modification of a particle propagator 
is assumed at very large spatial momentum: 
\be
\fr{i}{\omega^2 -{\bf k}^2 } \rightarrow \fr{i}{\omega^2  - \fr{{\bf k}^6}{\LambdaHL^4}}.
\label{HLprop}
\ee 
While such a propagator leads to better convergent loop
integrals, the absence of higher derivatives with respect to time 
in the Lagrangian, and consequently the absence of 
$\omega^4$ etc. terms in the propagator allows one to
extend the regime of validity of this theory beyond 
$\LambdaHL$ without immediately encountering 
pathological ghost-like features. But at the same time such a 
construction leads to the violation of Lorentz symmetry explicitly
above the Lifshitz scale. If, however, a theory of this
type is coupled to SM sector through power-suppressed interactions
only, it is conceivable that the size of induced LV  terms in SM is controlled
by the ratio $\LambdaHL^2/M^2$ and can be made small,
given a sufficiently large separation between $\LambdaHL$ and $M$. 
There would be no need of fine-tuning since radiative corrections
become stabilized so that $\LambdaHL \ll M$ alone would be sufficient.

We shall illustrate this mechanism in a toy example
with a neutral fermion that has a Lifshitz-type propagator.
It couples to photon through an anomalous magnetic moment,
which is a power-suppressed interaction.
In this case, as expected, the LV 
corrections induced by the fermion 
to the photon sector is controlled by 
$\mu^2 \LambdaHL^2$, where $\mu$ is the anomalous magnetic moment. 
Given that this product can be made arbitrarily small, approximate 
Lorentz symmetry in the photon sector is maintained
despite being completely broken for the neutral fermion.

Perhaps the most interesting example of this type would be gravity, since 
its interactions are suppressed by a very large scale.
Besides many interesting features of Lifshitz type field theories
that have been intensively studied in the past,
it has attracted a lot attention when 
it was proposed by Ho\v rava that a theory of this type stands as 
a candidate for a renormalizable 
theory of gravity \cite{Horava:2009uw}.
Among different issues that Ho\v rava's theory for gravity
is facing at phenomenological level, the question of LV is not 
the last on the list. Given that the graviton propagators 
violate Lorentz symmetry in the ultraviolet, is it reasonable 
to expect that such a theory would respect Lorentz symmetry at low energies
without tremendous fine-tuning? The answer to this 
question is by no means a straightforward one. 
If Ho\v rava-Lifshitz type behavior is more than just a 
cute way of making loops better convergent, but indeed a 
description of nature at 
short distances, one has to specify how this behavior is 
consistent with stringent tests of Lorentz symmetry performed 
with a variety of the SM particles. 
We have two classes of interaction:
those that have dimensionless couplings in the Standard Model
($\alpha_s$, $\alpha_W$, $\alpha_{EM})$, and gravity whose 
strength is controlled by Newton constant 
$G_N = \fr{1}{8\pi \MP^2}$. 
Various loop corrections to the propagation of SM particles 
will have different types of divergences, and
all of them  must not introduce an overwhelming amount of LV. 
A priori, one has the option
as to where to put Lifshitz behavior: in the matter sector, 
   in the gravity sector,
or in both. We shall distinguish two generic
options: 

\begin{itemize} 

\item {\em Option 1} Both SM and gravity sectors flow into the Lifshitz-type behaviour above $\LambdaHL$. 

\item {\em Option 2} Only gravitational propagators become Lifshitz-type at 
$\LambdaHL$, while the bare SM action preserves 
normal Lorentz symmetric propagators all the way to the Planck scale. 

\end{itemize}

Option 1 leads to fine-tuning issues even in the limit as gravity is decoupled.
Indeed, various SM loop corrections to the dim-4 
kinetic operators are not universal for different types 
of particles: {\em e.g.} compared to leptons and photons, quarks and gluons will have extra corrections due to the 
strong group  etc. {\em In the absence of additional protective symmetries}, this should lead to a Lorentz non-universality 
of radiative corrections. Even if one assumes an exact 
universality of the speed of propagation for different 
species, simple one-loop corrections would introduce a non-universality 
of the order of 
$\alpha_{SM}/\pi \sim 10^{-3}-10^{-2}$, which has to be 
tuned away at 1 part per $10^{20}$. This was recently 
illustrated by the calculation of 
radiative corrections in the toy model that 
involved two different scalar fields \cite{Iengo:2009ix}.
Therefore it seems that this option is troublesome even 
before the gravity effects are taken into account and regardless 
of whether one has a large scale separation between $\LambdaHL$ 
and Planck scale. 

Option 2 seems to be more viable. Indeed all the loop corrections that involve SM fields but not gravity are automatically
Lorentz-preserving. The fact that gravity couples to the 
SM fields only through Planck mass suppressed interaction
leads us to consider the protection mechanism outlined above.  
Our proposal is that the 
gravitational loops (which normally would be power-divergent)
get stabilized at Ho\v rava-Lifshitz scale so that possible 
non-universality generated through quantum corrections
in the propagation speed of different species, 
indicating that the induced LV in the SM sector scales as 
\be
\label{naive}
\Delta c \sim \fr{\LambdaHL^2}{ \pi^2 \MP^2}.
\ee
Similar to the toy example discussed earlier, 
one could hope to have a control over this quantity
via the ratio $\LambdaHL^2/\MP^2$, and, demanding
sufficient scale separation, ensure that it is small.

We perform a detailed one-loop analysis of Ho\v rava-type 
gravity, calculating corrections to the speed of propagation 
for vectors and scalars, and find that loop corrections
produced by the spin-2 and spin-1 graviton do indeed exhibit
the behavior described by (\ref{naive}), but
some quadratic divergences associated with the vector-graviton loop
diagrams remain. We see that these remaining quadratic
divergent corrections are not universal between scalars 
and vectors, thus potentially reinstating the
issue of fine-tuning in the theory.  

Our analysis, however, points
toward a relatively easy solution to the fine-tuning problem. 
The inclusion into the action of a single term that respects all
the symmetries of the original model of Ho\v rava-Lifshitz gravity 
but with a Lifshitz dimension higher than $6$, counted in
a naive way, is sufficient
to suppress all quadratically-divergent contributions and render 
the loop-induced Lorentz violation in the Standard Model sector completely
under control. In such an extended model, the mechanism we conjectured above
is fully at work, and the need for fine-tuning to maintain
the Lorentz symmetry in consistency with the observations is 
totally absent.  The model, on the other hand, might still 
harbor additional problems associated with the new terms we introduce, 
and we will defer extended discussion on this topic to the follow-up works.

The status of Ho\v rava's original proposal
\cite{Horava:2009uw} as well as its various extensions
\cite{Sotiriou:2009gy, Blas:2009qj, Horava:2010zj}, 
both on theoretical and phenomenological 
ground, is still being actively discussed and
debated in the literature \cite{Mukohyama:2009zs, Charmousis:2009tc, Li:2009bg,Sotiriou:2009bx,
	Blas:2009yd, Bogdanos:2009uj,Afshordi:2009tt,
		Henneaux:2009zb, Papazoglou:2009fj,
Blas:2009ck, Blas:2010hb, Padilla:2010ge, Mukohyama:2010xz, Kimpton:2010xi,Kobakhidze:2009zr, Orlando:2009en, Orlando:2009az}. 
We make no attempts to delve on these issues
 in the current study, concentrating instead on the 
perturbative calculation at
one-loop level using linearized gravity action 
to illustrate our main
points. Furthermore, to make calculations more straightforward we work
within the ``healthy extension'' framework proposed in \cite{Blas:2009qj},
and assume that the full non-linear theory is consistent  provided that the parameters are chosen
properly. It would become clear that our main conclusion is largely independent
of the specific choice of those model dependent
parameters. We mention in passing that
Lorentz violating effects in Ho\v rava's gravity, being considered
from very different angles, were also discussed in other works 
\cite{Chen:2009ka, Alexandre:2010aq, Alexandre:2010vh}.

Graviational loop calculations can be cumbersome, 
not least due to the necessity of introducing 
explicit gauge fixing in the gravity sector. At one-loop level,
quantum corrections to the effective
action for each individual particle is gauge-choice
dependent,  but fortunately, such dependence is always
canceled out when one compares the same correction for different matter fields.
The \emph{actual Lorentz violation effect} we present
in this paper, exhibited by dimension-$4$ operators, i.e.
the difference of the propagation speeds of massless particles with
different spins or  other quantum numbers, 
is independent of the gauge choice and therefore bares true physical
meaning.  Of course, it is only such differences that are
physical since the radiative corrections to the propagation
speed common for all matter fields can be easily absorbed by
a rescaling of space and time coordinates.

This paper is organized as follows. In the next section we analyze a 
toy model with a neutral Lifshitz-type 
fermion interacting with photon via the magnetic moment, and 
calculate its radiative corrections to the photon 
action and the induced Lorentz violation. 
In section 3 we introduce the Ho\v rava-Lifshitz type
theories for gravity, truncate the action to the 
quadratic level and derive the propagators for the gauge invariant modes. 
In section 4 we calculate
the difference of the propagation speed for vectors and scalars,
both minimally coupled to gravity, where we will find
residual fine-tuning in the standard Ho\v rava-Lifshitz models.
Section 5 presents a simple extension to the  same model where such
fine-tuning can be eliminated.  We include further 
discussion in Sec. 6.  More details regarding the loop calculations
are presented in Appendix \ref{app:propagators} and 
\ref{app:lifshitz_fermion}.  Appendix \ref{app:toys} 
includes two toy models for the Lifshitz type QED, which, being
gauge theories, share a lot  of common properties and issues with 
the Ho\v rava-Lifshitz gravity. 

A few words on the convention we would follow in this paper:
we will consider only one-loop diagrams,
which either consist
of only one propagator and one vertex, or two propagators and two vertices.
Each vertex contains a factor of $\frac{1}{i\hbar}$ and it cancels precisely the
factor of $i\hbar$ carried by each propagator. Consequently, we can
safely ignore these factors altogether.
Just to fix the notation, if the action is given by the form
$S=-\frac{1}{2}\phi\mathcal O\phi+\lambda\phi^2$, we would
say the propagator is $\mathcal O^{-1}$ and the vertex is $\lambda$.
We shall also use the convention that $\square=-\partial_t^2+\Delta$.
Its transformation into the momentum space is given by the rule
$\partial_t\rightarrow -i\omega$ and $\partial_i\rightarrow ik_i$, 
and therefore $\square\rightarrow \omega^2-\vec k^2$.

\section{A toy model of a neutral Lifshitz fermion}
\label{sec:toy}
\label{sec:fermion}

Let us first consider a simple toy example. Suppose we have a 
Lifshitz-type neutral fermion whose action is given by
\begin{equation}
\mathcal L_\psi=\bar\psi\left[
\gamma^0\partial_t+\LambdaHL^{1-z}\left(\sqrt{-\Delta}\right)^{z-1}
		\gamma^i\partial_i\right]\psi\,,
\end{equation}
where we have introduced a Lifshitz scale $\LambdaHL$ and 
the Lifshitz critical exponent $z$. 
When $z>1$, this action has an
anisotropic scaling behavior. In principle one should
include all the lower spatial derivative terms, but at
large $\vec k$, which is the limit that we are mainly interested in,
the highest spatial derivative term
dominates and we will keep only it. 

Let us suppose that this fermion couples to photon through an 
irrelevant operator given by
\begin{equation}
\label{magnmom}
\mathcal L_I=
\frac{1}{2M}F^{\mu\nu}\bar\psi\sigma_{\mu\nu}\psi\,,
\end{equation} 
where $M$ is a mass parameter which gives 
the fermion an anomalous magnetic moment $\mu = M^{-1}$.
The photon kinetic term takes the usual form 
$\mathcal L_A=-\frac{1}{4}F_{\mu\nu} F^{\mu\nu}$.

We would like to evaluate the fermion 1-loop correction
to the photon kinetic operator. In particular we are 
looking for Lorentz symmetry violating effect.
It is useful to define
\begin{equation}
(\tilde k^0,\, \tilde{\vec k})\equiv
(k^0, \,|k|^{z-1}\vec k/\LambdaHL^{z-1})
\end{equation}
and
\begin{equation}
\tilde k^2\equiv -k_0^2+\frac{|\vec k|^{2z}}{\LambdaHL^{2z-2}}\,.
\end{equation}
With these notations, the fermion propagator is
$1/\dslash{\tilde k}$.

The one-loop integral that contributes to the photon kinetic
operator in the zero external momentum limit is given by
\begin{equation}
K=-\frac{1}{ 4 i (2\pi)^4  M^2}\int \ud ^4 k \quad
\frac{F^{\mu\nu} F^{\alpha\beta}
	\tr\,\sigma_{\mu\nu}\dslash{\tilde k}\sigma_{\alpha\beta}
	\dslash{\tilde k}}{\tilde k^4}\,.
\end{equation}
Detailed calculation of this integral is presented 
in Appendix \ref{app:lifshitz_fermion}.
It is found that when $z=1$ and the theory respects
the Lorentz symmetry, $K$ vanishes identically leading to 
no correction to the photon kinetic term
at this level.  When $z>1$ and the Lorentz symmetry is broken,
\begin{equation}
K=\frac{\LambdaHL^{3(1-1/z)}(f^t-f^x)}{2 M^2} (\mathbf E^2+\mathbf B^2)\,,
\end{equation}
where
\begin{equation}
f^t\equiv -\frac{8}{i (2\pi)^4}
\int \ud^4k\, \frac{k_0^2}{z |\vec k|^{3(1-1/z)} k^4}\,,
\end{equation}
and
\begin{equation}
f^x\equiv \frac{8}{3i (2\pi)^4} \int \ud^4k\, \frac{|\vec k|^2}{z |\vec k|^{3(1-1/z)} k^4}\,.
\end{equation}
As $z=3$, both $f^t$ and $f^x$ are logarithmically divergent
and it turns out that $f^t=3f^x$. Consequently, including the 
one loop correction, the photon kinetic term becomes
\begin{equation}
\mathcal L_A=
\frac{1}{2}\left(1+\frac{\LambdaHL^2}{3\pi^2 M^2}
		\log\frac{\Lambda_{\textrm{UV}}}{\LambdaHL}\right)
\mathbf E^2
-\frac{1}{2}\left(1-\frac{\LambdaHL^2}{3\pi^2 M^2}\log\frac{\Lambda_{\textrm{UV}}}{\LambdaHL}\right)
\mathbf B^2
\end{equation}
which leads to an effective ``speed of light''
\begin{equation}
\label{scaling}
c'^2=\left(1-\frac{2\LambdaHL^2}{3\pi^2 M^2}
\log\frac{\Lambda_{\textrm{UV}}}{\LambdaHL}\right) c^2\,.
\end{equation}
One can easily see that this correction is 
under control if there is  a wide scale separation between $\LambdaHL$ and $M$. 
Notice also that the Lorentz symmetry of the interaction term in (\ref{magnmom}) is not 
essential for the scaling (\ref{scaling}) to hold. We could, for example, 
``disbalance'' $\sigma_{0i}F^{0i}$ and $\sigma_{ij}F^{ij}$ in  a LV way, which would affect the numerical coefficient
in (\ref{scaling}), but not change the ratios of the dimensionful scales.

\section{Action and propagators for the Ho\v rava type gravity}
\label{sec:gravity}
From this point on, we would like to consider quantum corrections to ordinary
matter fields coupled to a Ho\v rava-Lifshitz type gravity.
The main point is that since gravity is coupled to matter through
irrelevant couplings, the loop effects are suppressed by $1/\MP^2$, but
this suppression is compensated in GR by a quadratic UV divergence. 
Such divergences have been encountered in previous calculations of 
LV effects with graviton loops (see {\em e.g.} \cite{Burgess:2002tb,Withers:2009qg}).
Ho\v rava
gravity has the virtue that at least some parts of the loop diagrams are more
convergent since the graviton propagator scales anisotropically
at large momentum. For Lifshitz critical exponent $z=3$,
the better convergent
loops are logarithmically divergent only, leading to a 
logarithmic running of the effective
speed of light in the matter sector. 
If all the loop induced quantum corrections
are logarithmically divergent as such, it is
conceivable that given a wide separation between the 
scale $\LambdaHL$ and $\MP$,
similar to what we found in the previous section, the induced 
violation of the Lorentz symmetry might be under control.
The main physics question to be addressed is 
whether the matter actions acquire quantum corrections that
lead to the non-universality of the propagation speed, and if 
so with what coefficients. In fact, as
we would show below, such corrections are generically not universal and different
$c^2$ for different species induces observable LV effects.  

In this section, we briefly describe the gravity theory of interest.
The fact that Ho\v rava theory is a gauge theory,
which contains constraints and non-dynamical fields, makes the problem
more involved compared to the simple toy example presented above.
We will find that at one-loop level, the theory exhibits mixed
properties: while some loops are better convergent as expected, 
others remain quadratically divergent.
Non-linearity makes any gravitational theory
quite difficult to analyze perturbatively without running into various
subtleties. The physics is much more transparent in simpler examples
such as a Lifshitz-type quantum electrodynamics, which we present
in Appendix \ref{app:toys} as an analogy to the calculation 
we perform for the gravity case below.

We define the fields for the metric perturbations 
above the flat spacetime background as
\begin{gather}
-g_{00}=1+n\,,\\
g_{0i}=n_j\,,\\
g_{ij}=\delta_{ij}+h_{ij}\,.
\end{gather}
Einstein's theory of general relativity, expressed in ADM formalism,
is described by the Lagrangian 
$\mathcal L_{\textrm{EH}}=\MP^2\sqrt{\gamma}N(R+K_{ij} K^{ij}-K^2)$.
The action for Ho\v rava gravity is different 
from it in two aspects, both leading
to the violation of Lorentz symmetry. In the low momentum limit
it differs from GR in that the
combination $K_{ij}K^{ij}-K^2$ is replaced by a more general expression
$K_{ij} K^{ij} -\lambda K^2$, where a model-dependent parameter
$\lambda$ is introduced. 
In the large momentum limit, it is proposed
that higher spatial derivative terms dominate the action
and they are the key ingredients that render the graviton loop
more convergent and the theory renormalizable. For our
purpose, the highest
dimensional operators are the most important, and they include 
$R_{ij}\Delta R^{ij}$ and $R\Delta R$. 
We adopt the so-called ``healthy extension'' \cite{Blas:2009qj}
of the original theory in this paper, where additional terms such as
\begin{equation}
R\Delta^2 n=-\frac{2\sigma\Delta^3 n}{\MP^2}
\end{equation}
and $n \Delta^3 n$ are also needed to completely ``Lifshitzise'' the 
scalar sector.
All the fields introduced above are spacetime-dependent
functions and it is the so-called ``non-projectable'' version 
that is being considered here.
We parameterize the highest spatial derivative terms by
\begin{equation}
\mathcal L_{\textrm{Ho\v rava}}=\MP^2\left(\dots+\LambdaHL^{-4}R_{ij}\Delta R^{ij}
+\frac{a-3}{8}\LambdaHL^{-4}R\Delta R+\frac{b}{2}\LambdaHL^{-4}n\Delta^3 n
-\frac{c}{2}\LambdaHL^{-4}R\Delta^2 n\right).
\end{equation}
Here, a Lifshitz scale $\LambdaHL$ as well as
three model dependent parameters $a$, $b$ and $c$ are introduced.
We will leave these parameters completely undetermined 
(other than requiring $b\neq 0$) and simply assume that 
some reasonable choices of these parameters exist such that
the theory is free from instabilities or strong coupling issues.
%and will restrict ourselves to the case $b\ne 0$. 
%The latter turns
%out to be essential as far as our calculations are concerned.

To derive the propagators, we 
expand the metric perturbation into different modes
that do not mix with each other, and then invert the kinetic term
in each sector individually.  It is most natural in this
setup to decompose the fields into different spin sectors with respect to
the $3$-dimensional rotational symmetry.  
From that point of view, $n$ is a scalar and we define
\begin{gather}
n_i=n_i^\T+\partial_i\varphi,\\
\label{eq:def_h_modes}
h_{ij}=h_{ij}^\TT
+\left(\partial_i V^\T_j+\partial_j V^\T_i\right)
+\left(\delta_{ij}-\frac{\partial_i\partial_j}{\Delta}\right)\sigma
+\frac{\partial_i\partial_j}{\Delta}\tau\,,
\end{gather}
where notation $\TT$ and $\T$ denote traceless-transverse and transverse
conditions respectively.  
We have altogether one transverse-traceless tensor $h_{ij}^\TT$,
two transverse vectors $V_i^\T$ and $n_i^\T$, and four scalars
$n$, $\varphi$, $\sigma$, and $\tau$.

Expanding the action $\mathcal L_{\textrm{Ho\v rava}}$
in terms of these variables to quadratic order, we decompose 
the result into three independent sectors,
which are referred to as the spin-2, spin-1, and spin-0
parts of the action and denoted by $\mathcal L_\mathbf{2,1,0}$
respectively. Explicitly, they are
\begin{equation}
\begin{aligned}
\mathcal L_{\mathbf{2}}=&\frac{1}{4}\dot h^{\TT 2}_{ij}
+\frac{1}{4}h^{\TT ij}\Delta h^\TT_{ij}+
\frac{1}{4\Lambda^4_L}h_{ij}^\TT\Delta^3 h^{\TT ij}
\,,\\
\mathcal L_{\mathbf{1}}=&-\frac{1}{2} \left(\dot V^\T_i-n^{\T i}\right)
	\Delta\left(\dot V^{\T i}-n^\T_i\right)\,,\\
\mathcal L_{\mathbf{0}}
=&\frac{1-2\lambda}{2}\dot\sigma^2-\frac{1}{2}\sigma\Delta\sigma
-(\lambda-1)\left(\Delta \varphi-\frac{1}{2}\dot\tau\right)^2
	+\lambda\dot\sigma\left(2\Delta\varphi-\dot\tau\right)
-2n\Delta\sigma\\
&\quad+\frac{a}{2\Lambda^4_L}\sigma\Delta^3\sigma
+\frac{b}{2\LambdaHL^4}n\Delta^3 n+\frac{c}{\LambdaHL^4}\sigma\Delta^3 n\,.
\end{aligned}
\end{equation}

Since $\lambda$ appears only in $\mathcal L_{\mathbf{0}}$, both
$\mathcal L_{\mathbf{1}}$ and $\mathcal L_{\mathbf{2}}$ are identical 
for $\mathcal L_{\textrm{EH}}$ and $\mathcal L_{\textrm{Ho\v rava}}$
if higher derivative terms are omitted.

In a truncated expansion of the gravity
action the full diffeomorphism symmetry is lost but a 
``partial gauge symmetry'' remains.
It is easily verified that the action
given above makes explicit the following gauge symmetry:
\begin{gather}
\label{eq:vector_gauge}
V_i^\T\rightarrow V_i^\T+\zeta^\T_i\,,\quad
n^\T_i\rightarrow n^\T_i+\dot\zeta^\T_i\,,\\
\label{eq:scalar_gauge}
\varphi\rightarrow\varphi+\dot\omega\,,\quad
\tau\rightarrow\tau+2\Delta\omega\,,
\end{gather}
where $\zeta_i^\T$ and $\omega$ are arbitrary infinitesimal functions.
When $\lambda=1$, the linearized Einstein-Hilbert action enjoys an additional
gauge symmetry generated by
\begin{equation}
\label{eq:scalar_gauge_GR}
n\rightarrow n-2\dot \chi\,,\quad
\varphi\rightarrow \varphi+\chi\,.
\end{equation}

For future purposes, we also define the gauge invariant combinations:
\begin{equation}
\label{eq:gauge_invariant_fields}
v_i^\T\equiv \dot V_i^\T-n_i^\T\,,\quad
\chi\equiv \frac{1}{2}\dot\tau-\Delta\varphi\,,
\end{equation}
which, instead of $V^\T_i, n^\T_i, \tau$ and $\varphi$,
are the real ``gauge-independent degrees of freedom''.
In Ho\v rava-Lifshitz gravity,
the ``fourth'' gauge symmetry is missing, so that $n$ is ``physical'' by itself
(apart from the time reparameterization symmetry).

The action $\mathcal L_{\mathbf{2}}$ leads to the 
propagators for spin-2 gravitons without any gauge ambiguity.
Directly read off from the action, it is given by
\begin{equation}
\vev{h^\TT_{ij} h^\TT_{kl}}=
	-\frac{\Pi(\vec k)_{ij, kl}}
{\omega^2-\vec k^2-\LambdaHL^{-4}\vec k^6}\,,
\end{equation}
where
\begin{equation}
\label{eq:def_Pi}
\begin{split}
\Pi(\vec k)_{ij, kl}
=&\left(\delta_{ik}-\frac{k_i k_k}{\vec k^2}\right)
\left(\delta_{jl}-\frac{k_j k_l}{\vec k^2}\right)
+\left(\delta_{jk}-\frac{k_j k_k}{\vec k^2}\right)
\left(\delta_{il}-\frac{k_i k_l}{\vec k^2}\right)\\
&\qquad
-\left(\delta_{ij}-\frac{k_i k_j}{\vec k^2}\right)
\left(\delta_{kl}-\frac{k_k k_l}{\vec k^2}\right)\,.
\end{split}
\end{equation}

The propagators given by $\mathcal L_{\mathbf{1}\,,\mathbf{0}}$, 
to the contrary, cannot be determined without making a gauge choice.  
The technical details, including the gauge fixing and the propagators,
are presented in Appendix \ref{app:propagators}.
In what follows, as much as possible, we carry out 
our calculations without choosing any particular gauge and express
the results in terms of the physical quantities consisting of
gauge-invariant combinations only. It will be shown
that our final conclusion is valid in general and manifestly independent
of the gauge conditions. 
%We will only apply the gauge fixing and the
%explicit formulae for the propagators given in Appendix \ref{app:propagators}
%to such the final result so that the numbers extracted there are
%real physical quantities.

\section{Loop-induced Lorentz violation in the matter sector}
\label{sec:LV}
We will consider in this section one-loop corrections to
the matter kinetic terms due to their coupling to gravity
described by a Ho\v rava-type theory.
Our goal is to compute the radiative corrections 
to the effective propagation speed for different species.
Any difference, $c_{\rm species~1} - c_{\rm species~2} \neq 0$, 
if present, would indicate the violation of Lorentz
symmetry at the quantum level.

We briefly mention the strategy for the calculation. Since
we are only interested in the one-loop corrections, it is sufficient to
expand the action in  metric perturbations up to  quadratic order.
For those terms that are linear in metric perturbations,
we ``square'' them to form a one-loop diagram, using 
two vertices, each containing one graviton leg. For 
these diagrams, the loop is formed
by one graviton propagator and one matter propagator. For
terms quadratic in metric perturbation,
we must form a closed graviton loop with single graviton propagator.
We focus on the leading divergent contributions,
and therefore will set the external momentum to zero inside the loop integrals.
Moreover, we are interested only in those one-loop radiative corrections
to the matter kinetic term that can actually lead to violation of 
the Lorentz symmetry, and therefore, it suffices to 
expand $\sqrt{-g}$ to the first order 
because at one-loop level the radiative corrections
from the quadratic expansion of $\sqrt{-g}$
can only renormalize $\MP$.  Since terms that contain
metric perturbations at quadratic order contribute only
when the metric perturbations are contracted among themselves forming
a single graviton closed loop, we are allowed to
replace all the quadratic expression  
of the metric perturbations in the action
by their correlation functions directly, which entails
a sequence of simplifications.
For example, a term in the action of the form
$F_{ij}(h_{kl}, n, n_k)\partial^i\phi\partial^j\phi$,
where $F_{ij}$ is a \emph{quadratic} expression of the metric perturbations,
can be equivalently replaced by its correlation function
$\vev{F_{ij}}=\frac{1}{3}\vev{F_{kl}\delta^{kl}}\delta_{ij}$.
In this last step, we have made use of the three-dimensional rotational symmetry 
that remains valid in Ho\v rava's gravity.
Similarly, terms of the form of
$F_i(h_{kl}, n, n_k)\partial^i\phi\dot\phi$
where $F_i$ is a three dimensional vector that is also a \emph{quadratic} function
of the metric perturbations cannot contribute at 1-loop level and will
be omitted.  Vertices that mix different spin components of gravitons
do not contribute at one loop either
and are omitted.  We will apply these simplifications implicitly 
from this point on, and, for brevity, drop the $\vev{\cdot}$ sign in the action if
any quadratic expression of the metric perturbation fields 
are replaced by the corresponding correlation function.  

We would repeatedly encounter the divergent integrals:
\begin{equation}
\mathbb L\equiv \frac{1}{i(2 \pi)^{4}}
  \int\frac{\ud\omega\ud^3 k}{\omega^2-\LambdaHL^{-4}\vec k^6}\,,
\end{equation}
and when the fields are canonically normalized and the proper scales 
are restored, we have
$\mathbb L= \frac{\LambdaHL^2}{8\pi^2\MP^2}
\log\frac{\LambdaUV^2}{\LambdaHL^2}$.

Finally, let us fix the convention for the notation of
correlation functions.
Since all correlation functions considered here are two-point
functions of two operators, and we always take the external momentum to 
zero, we can omit the ``,'' and denote $\vev{AB}\equiv \vev{A\,,B}$,
which, in momentum space, should be understood as
$ \vev{A(\omega, \vec k)\,,B(-\omega, -\vec k)}$.
For brevity, we also introduce a notation for 
the correlation functions of two identical operators: we 
denote $\twopt{A}\equiv\vev{A(\omega,\vec k) A(-\omega, -\vec k)}$.

Let us first consider a scalar $\phi$ minimally coupled to gravity,
described by the Lorentz-symmetric ``bare'' Lagrangian 
\begin{equation}
\begin{split}
\mathcal L=&-\frac{1}{2}\sqrt{-g} g^{\mu\nu}\partial_\mu\phi\partial_\nu\phi\,,
\end{split}
\end{equation}
whose propagator is of course just $\vev{\phi\,\phi}=-1/(\omega^2-\vec k^2)$.
It is meant to represent an elementary SM matter field, such as {\em e.g.} the Higgs field.

Following the strategy explained above, we expand the action to
the quadratic order in terms of the metric perturbations and decompose
the interaction terms into each spin sector defined by the relevant
metric perturbations involved. Explicitly, we have
\begin{equation}
\mathcal L^\textrm{I}_{\mathbf 2}=
\left(\frac{1}{2}h^{\TT ij}-\frac{1}{2\cdot 3}
h^{\TT kl} h_{kl}^\TT\delta^{ij}\right)
\partial_i\phi\partial_j\phi\,,
\end{equation}
\begin{equation}
\mathcal L^\textrm{I}_{\mathbf 1}=
-n^{\T i}\partial_i\phi\dot\phi
+\partial^i V^{\T j}\partial_i\phi\partial_j\phi
-\frac{1}{3}\left(
\partial_k V^\T_l \partial^k V^{\T l}+n^\T_k n^{\T k}\right)
\partial^i\phi\partial_i\phi
+\dots\,,
\end{equation}
and
\begin{equation}
\begin{aligned}
\mathcal L^\textrm{I}_{\mathbf 0}=&
\frac{1}{2}\left(\sigma+\frac{1}{2}\tau-n\right)\dot\phi^2
-\partial^i\varphi\partial_i\phi\dot\phi
+\frac{1}{2}\left[\frac{\partial_i\partial_j}{\Delta}(\tau-\sigma)
	-\frac{1}{2}\tau\delta^{ij}-n\delta^{ij}\right]
\partial^i\phi\partial^j\phi\\
&+\frac{1}{2}(2n^2)\dot\phi^2
-\frac{1}{2}\cdot\frac{2}{3}\left[n(2\sigma+\tau)-\sigma\tau
+\frac{1}{4}\tau^2+\partial^k\varphi\partial_k\varphi\right]
\partial^i\phi\partial_i\phi+\dots\,.
\end{aligned}
\end{equation}
Here, ellipses stands for terms that are manifestly Lorentz
invariant, which we drop in the subsequent calculations.

Let us denote the 1-loop radiative correction to the kinetic
term of $\phi$ as 
\begin{equation}
\delta(\partial_\mu\phi\partial^\mu\phi)=
\frac{1}{2}(K^t\dot\phi^2-K^x\partial^i\phi\partial_i \phi)\,,
\end{equation} 
and the contributions from each part of the interaction
$\mathcal L^\textrm{I}_\mathbf{2,1,0}$ as 
$K^t_{\mathbf{2}, \mathbf{1}, \mathbf{0}}$ and
$K^x_{\mathbf{2}, \mathbf{1}, \mathbf{0}}$ respectively.

It is clear that $K^t_\mathbf{2}=0$ and
$K^x_\mathbf{2}=-\frac{4}{3}\cdot\mathbb{L}$.
The contribution induced by the vector-gravitons is also easy to compute
and the result is identical to that in GR.
We find $K^t_{\mathbf{1}}=0$, and \footnote{It so happens that in the current theory
\begin{equation}
\int\ud\omega\ud^3 k \vev{\partial_i V_j^\T \partial^i V^{\T j}
	+n^\T_i n^{\T i}}=0\,,
\end{equation}
in any gauge when the symmetry-preserving 
regularization of the UV divergence is employed. 
In any event,  this term cancels out in the final answer by itself without
employing the vanishing of this loop integral.} 
\begin{equation}
\label{eq:k1x_scalar}
K^x_\mathbf{1}=
\frac{1}{3}\int \ud\omega\ud^3\vec k\,
\frac{\twopt{\dot n^{\T i}-\Delta V^{\T i}}}{\omega^2-\vec k^2}
+\frac{2}{3}\int\ud\omega\ud^3 \vec k\,
	\vev{\partial_k V^\T_l\partial^kV^{\T l}+n^\T_k n^{\T k}}\,.
\end{equation}
We denote the combination
given above as $K^x_{\mathbf{1}\,\textrm{scalar}}$. 
This expression is gauge choice
dependent and therefore cannot be physical. It will be cancelled
in the final result by other contributions, as we will show shortly. 

Similarly, we can compute the contributions from the spin-$0$ sector
as
\begin{equation}
\begin{split}
K^t_{\mathbf{0}}=&\int\ud\omega\ud^3\vec k\,
\left(-\frac{\twopt{\dot \sigma-\dot n+\chi}}
{\omega^2-\vec k^2}+\vev{2 n n}\right)\\
\end{split}
\end{equation}
and
\footnote{We attempt to express everything in terms 
of the gauge invariant combinations, in this case, $\chi$ as defined in 
\eqref{eq:gauge_invariant_fields}. To do so, identities as
$\frac{\vec k^2\omega^2}{\omega^2-\vec k^2}
=\left(\vec k^2+\frac{\vec k^4}{\omega^2-\vec k^2}\right)$
and $\frac{\vec k^2}{\omega^2-\vec k^2}
=\left(-1+\frac{\omega^2}{\omega^2-\vec k^2}\right)$ are used
so that we can trade time-derivative for spatial derivatives and
vice versa, at the cost of generating extra terms that can be
combined with those generated by the single graviton loop diagrams.
We will apply the similar identities while computing the spin-0 graviton
loop corrections to the photon kinetic term as well.}
\begin{equation}
\begin{split}
K^x_{\mathbf{0}}
=&\int\ud\omega\ud^3\vec k\,\left[
\frac{1}{3}\cdot\frac{\twopt{\chi-\dot\sigma-\dot n}}{\omega^2-\vec k^2}
+\vev{\frac{1}{12}\tau^2-\frac{1}{3}(\sigma-n)^2-\frac{1}{3}\sigma\tau
	+\tau n+\vec k^2\varphi^2}\right]\,.
\end{split}
\end{equation}

All contributions combined, we find that the effective change 
of the propagation speed for a neutral scalar, 
given by the difference between $K^x$ and $K^t$, is
\begin{equation}
\begin{split}
\delta c^2_{\textrm{scalar}}=&
-\frac{4}{3}\cdot \mathbb{L}+K^x_{\mathbf{1}\,\textrm{scalar}}\\
&+\int\ud\omega\ud^3\vec k\;\frac{1}{\omega^2-\vec k^2}
\left[
\frac{1}{3}\twopt{\chi-\dot\sigma-\dot n}
+\twopt{\dot \sigma-\dot n+\chi}\right]\\
&+\int\ud\omega\ud^3\vec k\,
\vev{\frac{1}{12}\tau^2-\frac{1}{3}(\sigma-n)^2-\frac{1}{3}\sigma\tau
	+\tau n+\vec k^2\varphi^2-2n^2}\,.
\end{split}
\label{scalarc}
\end{equation}

Let us do the same calculation for a $U(1)$ gauge field coupled to gravity.
The action for the minimally coupled photon is given by
\begin{equation}
\mathcal L=-\frac{1}{4}\sqrt{-g}g^{\mu\alpha}
g^{\nu\beta}F_{\mu\nu} F_{\alpha\beta}\,.
\end{equation}
To avoid choosing a gauge for photon, we work with the
physical fields $E_i=F_{0i}$ and 
$B_i=\frac{1}{2}\varepsilon_{ijk}F^{jk}$ and their correlation
functions.  It is easily verified that in any gauge
\begin{equation}
\vev{E_i\, E_j}=\frac{-\omega^2\delta_{ij}+k_i k_j}{\omega^2-\vec k^2}\,,\quad
\vev{B_i\, B_j}=\frac{-\vec k^2\delta_{ij}+k_i k_j}{\omega^2-\vec k^2}\,,\quad
\vev{E_i\, B_j}=\frac{\varepsilon_{ijn}\omega k^n}{\omega^2-\vec k^2}\,.
\end{equation}

Following the same procedure as before, we find the
relevant part  of the graviton-photon interactions,
separated into different spin sectors, 
is given by
\begin{equation}
\mathcal L^\textrm{I}_\mathbf{2}=
-\frac{1}{2}h^{\TT ij}\left(E_i E_j+B_i B_j\right)
-\frac{1}{2\cdot 6}h^{\TT ij}h^\TT_{ij} B^2+\dots\,,
\end{equation}
\begin{equation}
\begin{aligned}
\mathcal L^\textrm{I}_\mathbf{1}=&
-\partial_i V^\T_j (E^i E^j+B^i B^j)
	-\varepsilon_{ijk}\, n^{\T i} E^j B^k
-\frac{1}{6}\left(n^{\T i} n^\T_i+\partial_i V^\T_j \partial^i V^{\T j}\right)B^2
	+\dots\,,
\end{aligned}
\end{equation}
\begin{equation}
\begin{aligned}
\mathcal L^\textrm{I}_\mathbf{0}=&
-\frac{1}{2}\left[\left(n-\frac{1}{2}\tau\right)\delta^{ij}
+\frac{\partial^i\partial^j}{\Delta}(\tau-\sigma)\right](E_i E_j+B_i B_j)
-\varepsilon_{ijk}\partial^i\varphi E^j B^k\\
&+\frac{1}{2} (2n^2) E^2-\frac{1}{2}\left(\frac{1}{3}\sigma^2+\frac{1}{6}\tau^2
		+\frac{1}{3}\partial^i\varphi\partial_i\varphi\right) B^2
+\dots\,,
\end{aligned}
\end{equation}
and, again, ``$\dots$'' represents those terms shared by both $E^2$ and $B^2$
that do not lead to any Lorentz symmetry violation.

Let us consider the 1-loop correction to the photon kinetic term, 
denoted similarly as
$\frac{1}{2}(K^t E^2 -K^x B^2)$. Again, the contributions
from each spin sector $\mathcal L^\textrm{I}_\mathbf{2,1,0}$ 
are denoted as $K^t_{\mathbf{2}, \mathbf{1}, \mathbf{0}}$
and
$K^x_{\mathbf{2}, \mathbf{1}, \mathbf{0}}$ respectively.
It is most easily checked that
\begin{equation}
K^t_\mathbf{2}=\frac{4}{3}\cdot\mathbb L\,,\quad\textrm{and}\quad
K^x_\mathbf{2}=-\frac{2}{3}\cdot\mathbb L\,.
\end{equation}
To evaluate the contributions from the spin-1 sector, the vertex
$\varepsilon_{ijk}n^{\T i} E^j B^k$ is very important. When all
the crossing terms are properly included, one finds that many terms combine
into a ``complete square'' so that $K^t_{\mathbf{1}}$ can be
expressed as
\begin{equation}
\begin{aligned}
K^t_\mathbf{1}=&-\frac{1}{3}\int\ud\omega\ud^3 k \vec k^2
\vev{V^\T_i V^{\T i}}
-\frac{1}{3}\int\ud\omega\ud^3 k\,
\frac{\vec k^2\vev{v^\T_i \, v^{\T i}}}{\omega^2-\vec k^2}\,,
\end{aligned}
\end{equation}
where $v^\T_i=\dot V^\T_i-n^\T_i$ is the gauge invariant combination
defined in Eq. \eqref{eq:gauge_invariant_fields}.
Similarly,
\begin{equation}
\begin{aligned}
K^x_\mathbf{1}=&
\frac{1}{3}\int\ud\omega\ud^3 k \vev{n_i^\T n^{\T i}}
+\frac{1}{3}\int\ud\omega\ud^3 k\, 
\frac{\twopt{\dot n_i^\T-\Delta V^\T_i}}
{\omega^2-\vec k^2}\\
&+\frac{1}{3}\int\ud\omega\ud^3 \vec k\,
	\vev{\partial_k V^\T_l\partial^kV^{\T l}+n^\T_k n^{\T k}}\,.
\end{aligned}
\end{equation}

The contributions from loop diagrams with a scalar-graviton
propagator are somewhat more cumbersome, but still straightforward to calculate.
With all the crossing terms included and everything expressed
in terms of the gauge invariant combinations whenever possible,
one eventually finds that
\begin{equation}
\begin{aligned}
K^t_\mathbf{0}=&\int\ud\omega\ud^3k\,\left[-\frac{2}{3}
\frac{\twopt{\dot n-\chi}}{\omega^2-\vec k^2}
-\frac{1}{3}\twopt{n+\frac{1}{2}\tau-\sigma}+\vev{2 n n}\right]\,,
\end{aligned}
\end{equation}
and
\begin{equation}
\begin{aligned}
K^x_\mathbf{0}=&\int\ud\omega\ud^3k\,
\left[\frac{2}{3}\frac{\twopt{\dot n-\chi}}{\omega^2-\vec k^2}
+\vev{-\frac{2}{3}\left(n-\frac{1}{2}\tau\right)^2+\frac{1}{3}\sigma^2+\frac{1}{6}\tau^2
+\vec k^2\varphi^2}\right]\,.
\end{aligned}
\end{equation}
Putting these formulae together, we find the effective change of the
speed of light for photon is
\begin{equation}
\begin{split}
\delta c^2_{\textrm{photon}}=&
-2\cdot\mathbb{L}+K^x_{\mathbf{1}\, \textrm{scalar}}
+\frac{1}{3}\int\ud\omega\ud^3 k\, 
\frac{\vec k^2\vev{v^\T_i \, v^{\T i}}}{\omega^2-\vec k^2}
+\frac{4}{3}\int\ud\omega\ud^3 k
\frac{\twopt{\dot n-\chi}}{\omega^2-\vec k^2}\\
&+\int\ud\omega\ud^3 k\,
\vev{-\frac{1}{3}n^2+\frac{1}{12}\tau^2+\frac{2}{3}\sigma^2
	+n\tau-\frac{2}{3} n\sigma-\frac{1}{3}\tau\sigma+\vec k^2\varphi^2-2n^2}\,,
\end{split}
\label{photonc}
\end{equation}
where $K^x_{\mathbf{1}\,\textrm{scalar}}$ is the exact
combination given in Eq. \eqref{eq:k1x_scalar}.

Now, we are are ready to 
examine the real Lorentz symmetry violating effect given by 
the difference of the graviton 1-loop correction 
to the propagation speed for different species,
e.g. scalar and photon field in the current case.
 The final answer, being the difference between Eqs. (\ref{photonc}) and
(\ref{scalarc}), is rather simple and it reads
\begin{equation}
\begin{split}
\delta c^2_{\textrm{photon}}-\delta c^2_{\textrm{scalar}}
=&-\frac{2}{3}\cdot\mathbb{L}+\frac{1}{3}\int\ud\omega\ud^3 k
\, \frac{\vec k^2\vev{v_i^\T v^{\T i}}}{\omega^2-\vec k^2}\\
&+\frac{4}{3}\int\ud\omega\ud^3 k\, \frac{\vec k^2\vev{\sigma n}
	-\vev{\chi\dot\sigma}}{\omega^2-\vec k^2}
-\frac{1}{3}\int\ud\omega\ud^3 k\, \frac{\omega^2+3\vec k^2}{\omega^2-\vec k^2}
\vev{\sigma^2}\,.
\end{split}
\end{equation}
It is this quantity that measures the \emph{actual violation 
of the Lorentz symmetry}, which cannot be simply scaled away
by field and coordinate redefinitions. 
In this final result, all gauge dependent quantities, including
$K^x_{\mathbf{1}\,\textrm{scalar}}$ and any correlation
functions that explicitly contain $\tau$ and $\varphi$, disappear.  
Therefore, it is fully physical and independent of the gauge
fixing scheme. The second term above is quadratically divergent,
leading to a residual fine-tuning problem in this model as we
discuss further below.  This divergence is the direct consequence of the 
non-Lifshitz behavior of propagators for the spin-1 gravitons. 
The second line, generated by the spin-0 
gravitons, on the other hand, leads only to logarithmic divergence, which can be 
easily seen from the explicit propagators given in Appendix \ref{app:propagators}.
All the model-dependent quadratic divergences contributed by the spin-0 gravitons
are completely cancelled out in this final answer, so that
the only remaining quadratic divergences comes from the vector-graviton
loops.

We can simplify this formula slightly further if we use the knowledge
that all propagators $\vev{\chi\dot\sigma}$, $\vev{\sigma\sigma}$,
and $\vev{\sigma n}$ are of Lifshitz type, 
and loop integrals can be reduced to 
\begin{equation}
\begin{split}
\int&\frac{\ud\omega\ud^3 k\, \omega^2}{\omega^2-\vec k^2}
\vev{\textrm{Lifshitz}}\approx\int\ud\omega\ud^3 k\, 
\vev{\textrm{Lifshitz}}+\textrm{finite terms,}\\
\int&\frac{\ud\omega\ud^3 k\, \vec k^2}{\omega^2-\vec k^2}
\vev{\textrm{Lifshitz}}=\textrm{finite}.
\end{split}
\end{equation}
Dropping all finite contributions, we have
\begin{equation}
\label{eq:final}
\begin{split}
\delta c^2_{\textrm{photon}}-\delta c^2_{\textrm{scalar}}
=&-\frac{2}{3}\cdot\mathbb{L}+\frac{1}{3}\int\ud\omega\ud^3 k
\,\left[\frac{\vec k^2\vev{v_i^\T v^{\T i}}}{\omega^2-\vec k^2}
-\left(\frac{4}{\omega^2}\vev{\chi\dot\sigma}+\vev{\sigma\sigma}\right)
	\right].
\end{split}
\end{equation}

Substituting in the explicit forms for the propagators given in Appendix
\ref{app:propagators}, we reach our final result in the current version of Ho\v rava-Lifshitz gravity:
\begin{equation}
\label{old_result}
(\delta c^2)_\textrm{photon}-(\delta c^2)_\textrm{scalar}
=-\frac{\LambdaHL^2}{12\pi^2\MP^2}\log\frac{\LambdaUV^2}{\LambdaHL^2}
-\frac{3\lambda+1}{(3\lambda-1)}\,
\frac{\LambdaHL'^2}{24\pi^2\MP^2}\log\frac{\LambdaUV^2}{\LambdaHL'^2}
-\frac{\LambdaUV^2}{24\pi^2\MP^2}\,.
\end{equation}
Here $\LambdaHL'$ is the model-dependent  Lifshitz energy scale defined in
Eq. \eqref{eq:lambdahlprime}.

Very similar results are found in the case of simple Lifshitz Abelian gauge
theory, which we demonstrate in Appendix \ref{app:toys}.

We will discuss the implications of this result and propose
ways to improve the model in order to eliminate all the quadratic divergence
in the next section.

\section{An improved model and the absence of fine-tuning}
\label{sec:improved}
Our calculations in the previous section 
show that Ho\v rava-Lifshitz gravity and its extensions discussed in the literature thus far 
induce Lorentz violation effects in the Standard Model sector with quadratic 
sensitivity to the cutoff. This poses serious problem since
the model has essentially no natural protection
against large Lorentz violation in the matter sector,
and therefore tremendous amount of fine-tunning is required to keep
the model consistent with observations. 
This quadratic divergence in 
$\delta c^2_{\textrm{photon}}-\delta c^2_{\textrm{scalar}}$
means that our proposal based on a large scale separation
$\LambdaHL/\MP \ll 1$ to protect the Lorentz symmetry
in the Standard Model does not work, and we 
must modify the theory in order to remove such remaining divergence. 

Given our formula \eqref{eq:final}, the problematic piece is easy
to spot. It is the vector-graviton contribution, identical
to those in GR, that leads to the problem because
\begin{equation}
\vev{v_i^\T\, v^{\T i}}=-\frac{2}{\vec k^2}\,,
\end{equation}
and does not go to zero at 
large $|\vec{k}|$ the same way the Lifshitz propagators do.
This part of the calculation entirely parallels its counterpart
in the Einstein's theory and therefore it is not at all surprising that
it remains quadratically divergent.

There are ways to modify the theory to remove the quadratic divergence. 
Naturally, one thinks of including in the theory a term that contains
$v_i^\T \Delta^2 v^{\T i}$ so that at large momenta the propagator
receives Lifshitz scaling, $ {v_i^\T\, v^{\T i}}\sim 1/\vec k^4$, 
sufficient to suppress the relevant
loop integral and make it logarithmic.  In the three-dimensional covariant notation,
such terms may originate either from $K_{ij}\Delta K^{ij}$ or $\nabla^i K_{ij}\nabla^k K^{kj}$. 
Both possibilities are usually not considered since their Lifshitz dimensions 
are higher than $6$ in the naive counting method.  
Note, however, such counting is 
questionable in theories with mixed Lifshitz and non-Lifshitz behavior 
considered in this paper.

The consequences of $K_{ij}\Delta K^{ij}$ or $\nabla^i K_{ij}\nabla^k K^{kj}$
terms in the action are not explored. One potential worry is the 
modification to the ordinary kinetic term 
for the spin-2 gravitons by $K_{ij}\Delta K^{ij}$ term, and to avoid this 
we shall consider the addition to the Ho\v rava-Lifshitz Lagrangian given by
\begin{equation}
\label{eq:kDk}
\mathcal L'=\frac{2}{\Lambda^2}\nabla^i K_{ij}\nabla^k K^{kj},
\end{equation}
so that at the linearized level it only modifies the spin-1 and spin-0 
graviton actions, and produces terms
\begin{equation}
\mathcal L' =\frac{1}{2\Lambda^2}v_i^\T\Delta^2 v^{\T i}-
\frac{2}{\Lambda^2}\chi\Delta\chi\,.
\end{equation}
We can easily repeat our calculation in this new model when such
terms are included. The propagators are given 
in Appendix \ref{app:propagators}, and using them
we find
\begin{equation}
\label{eq:new_result}
(\delta c^2)_\textrm{photon}-(\delta c^2)_\textrm{scalar}
=-\frac{\LambdaHL^2}{12\pi^2\MP^2}\left(1
		+\frac{\sqrt{(1-2\lambda)\alpha^{-1}}}{2(2\lambda-1)}\right)
\log\frac{\LambdaUV^2}{\LambdaHL^2}
-\frac{\Lambda^2}{12\pi^2\MP^2}
\log\frac{\LambdaUV^2}{\Lambda^2}\,.
\end{equation}
This expression contains logarithmically divergent pieces only, and we note 
that each of the spin-2, spin-1, and spin-0 sector contributes
one term. 

In the new theory with the additional term \eqref{eq:kDk}
included in the Lagrangian, the mechanism we proposed in the introduction
is fully at work.  One can safely 
put both $\LambdaHL$  and $\Lambda$ well below the Planck scale, and the entire
framework, consisting of both a Lifshitz type gravity and a nearly Lorentz invariant
Standard Model sector, would stay completely consistent with observations.

%%%%%%%%%%%%%%%%%%%%%%%%%%%%%%%%%%%%%%%%%%%%%%%%%%%%%%%%%%%%%%%%%%%%%%%%%%%%%%%%
%%%%%%%%%%%%%%%%%%%%%%%%%%%%%%%%%%%%%%%%%%%%%%%%%%%%%%%%%%%%%%%%%%%%%%%%%%%%%%%%
%                                  CONCLUSION
%%%%%%%%%%%%%%%%%%%%%%%%%%%%%%%%%%%%%%%%%%%%%%%%%%%%%%%%%%%%%%%%%%%%%%%%%%%%%%%%
%%%%%%%%%%%%%%%%%%%%%%%%%%%%%%%%%%%%%%%%%%%%%%%%%%%%%%%%%%%%%%%%%%%%%%%%%%%%%%%%
\section{Discussion}
\label{sec:discussion}
In this paper we  argue that  large amount of 
Lorentz violation in the irrelevantly coupled 
sectors (axions, gravity etc) can co-exist with the Lorentz-symmetric 
phenomenology of SM particles and fields,
provided that quantum corrections are stabilized by 
a Lifshitz-type behavior above
$\LambdaHL$, a scale that can be adjusted. 
This idea is of particular interest if the
LV sector is gravity and is described by a Ho\v rava-type theory.
The key to this proposal is the 
``self-regulating'' behavior of Lifshitz-type propagators that
participate in the loops. 
Given that one could entertain a possibility of very large 
energy scale separation, 
$\LambdaHL \ll \MP$, the induced differences in the speed of propagation 
for different SM species can be under control by the 
ratio $(\LambdaHL/\MP)^2$ and no fine-tuned choice of
bare parameters to maintain Lorentz symmetry will be needed.

Our explicit calculations for a generalized 
Ho\v rava type gravity coupled with conventional
matter fields have confirmed this expectation
in the following sense: those fields in the gravitational sector 
that fully acquire the anisotropic scaling, such as the truly
dynamical transverse and traceless gravitons, 
induce Lorentz violation controlled by  $(\LambdaHL/\MP)^2\log\Lambda_{\rm UV}$. 
The quadratic divergence of graviton loop is explicitly softened to the 
logarithmic one above the Ho\v rava-Lifshitz scale. 
However, our result, Eq. (\ref{old_result}), shows
that in the conventional extensions of Ho\v rava-Lifshitz gravity,
loop-induced Lorentz violating effects do contain a residual
quadratic divergence. This divergence is generated by the 
non-Lifhsitz parts of the gravitational action for the
vector-gravitons.
Therefore, for the choice of $\LambdaUV \sim \MP$ our 
idea of putting dimension 4 
LV operators under  control of a small ratio of two 
dimensionful parameters does not quite work there: LV from the 
Ho\v rava gravity sector will be efficiently transmitted to the SM sector 
with the quadratic sensitivity to the cutoff, $(\Lambda_{\rm UV}/\MP)^2$.  
(In some sense, the situation is reminiscent of non-commutative field theories, where certain divergences are self-regulated while others remain.)

Could this problem be resolved? 
A quick remedy we proposed here is to include terms that suppress
the vector-graviton propagator in the UV, such 
as $\nabla^i K_{ij}\nabla_k K^{kj}$. 
This addition term in the action ensures the Lifshitz-type behavior for the 
vector modes of the metric perturbations and is consistent with all the 
symmetries of Ho\v rava gravity. It contains no more than 
two time derivatives either as required.
Typically such terms are not considered
because of their higher Lifshitz dimension, but
in the theory with mixed behavior 
(Lifshitz for gravity and non-Lifshitz for matter)
the naive counting of scaling dimensions can be misleading. 
We conclude that
such terms appear to be admissible, essentially leading to the 
same softening of the gravitational loops integrals
of the spin-1 modes as the rest of the graviton field. 
With these terms included, the one-loop corrections to the propagation speed 
of different species are fully under control,
always proportional to $(\LambdaHL/\MP)^2\log\Lambda_{\rm UV}$.
Provided that
$\LambdaHL\ll\MP$, the induced Lorentz violation
effect in the Standard Model can be minimized to 
the phenomenologically acceptable 
level.  We reserve more detailed analysis of 
the proposed extension for the follow-up works.

We close up with additional comments on the viability of the whole setup, 
and various phenomenological options. 
 \begin{itemize}
 
\item  {\em On the choice of scale for $\LambdaHL$}. Our answer suggests the {\em maximum} 
scale for the transition to Ho\v rava-Lifshitz behavior. Given that various phenomenological constraints 
on dimension 4 LV operators are more stringent than $10^{-20}$, one would need to have 
$\LambdaHL \la 10^{10}$ GeV. This is an intemediate scale often appearing in particle physics, 
a geometric mean of weak and Planck scales. 
A much more definitive statement about the limit on $\LambdaHL$ can be made once we extend  
our calculations to actual SM fermions (electrons, quarks), which we plan to address in the future. 
On the other hand, nothing prevents choosing much lower scales for $\LambdaHL$ such as a TeV or even 
meV scales. The latter is the absolute minimum set by precision tests of gravity at sub-mm scales.

 \item  {\em Graviton propagation speed.}  So far we have considered only corrections to the 
propagation speed of matter, but graviton propagation is also of phenomenological interest. 
Deep inside the Ho\v rava-Lifshitz regime the gravitons are super-luminal
and therefore cannot be constrained by {\em e.g.} Cerenkov radiation. However, there are also 
much milder 1\%-level accuracy constraints on $c_{\rm graviton}$ coming from the gravitational 
energy loss of binary pulsars. There are no good arguments in this theory why the matter and 
gravity should propagate with the same speed in the IR, and possibly some additional 
emergent symmetry is required.

\item {\em Higher-dimensional operators and higher loop corrections.}
So far in our considerations we neglected external momenta of particles. 
This corresponds to explicitly calculating dimension 4 LV operators, while neglecting 
dimension 6. It turns out that the highest energy cosmic rays can also 
be (barely) sensitive to the Planck-scale normalized dimension 6 operators \cite{Gagnon:2004xh}. 
Investigating the actual size of these operators induced by graviton loops 
is worth of a separate investigation. Similarly, an important subject to address 
is the higher-loop order where normal SM radiative corrections and gravitational 
corrections are combined.

\item {\em Ho\v rava gravity and supersymmetry.}
Supersymmetry of the Ho\v rava type theories was considered recently in {\em e.g.} \cite{Xue:2010ih}. 
It may bring additional benefits of making the speed of light universal not only among matter fields 
but also for gravitons. In addition, the non-linear terms in the gravity action 
may be used as a way of breaking supersymmetry \cite{Petr:talk}, in which case 
on should expect the soft-breaking mass in the matter sector to scale 
as $m_{\rm soft} \sim \LambdaHL^2/\MP$. This is again suggestive of the 
intermediate scale of $10^{10}$ GeV as a reasonable choice for $\LambdaHL$.

\end{itemize}

%%%%%%%%%%%%%%%%%%%%%%%%%%%%%%%%%%%%%%%%%%%%%%%%%%%%%%%%%%%%%%%%%%%%%%%%%%%%%%%%
%%%%%%%%%%%%%%%%%%%%%%%%%%%%%%%%%%%%%%%%%%%%%%%%%%%%%%%%%%%%%%%%%%%%%%%%%%%%%%%%
%                               ACKNOWLEDGMENTS
%%%%%%%%%%%%%%%%%%%%%%%%%%%%%%%%%%%%%%%%%%%%%%%%%%%%%%%%%%%%%%%%%%%%%%%%%%%%%%%%
%%%%%%%%%%%%%%%%%%%%%%%%%%%%%%%%%%%%%%%%%%%%%%%%%%%%%%%%%%%%%%%%%%%%%%%%%%%%%%%%
\section*{Acknowledgments}
The authors would like to acknowledge useful discussions with N. Afshordi, T. Jacobson, 
L. Leblond, M. Serone and M. Trott. MP would also like to acknowledge illuminating discussions with 
S. Groot Nibbelink, many years ago and  long before this paper, 
on the improvement of UV loop behavior due to 
re-summed LV propagators. YS would like to thank O. Pujolas, S. Sibiryakov,
G. Gabadadze, and D. Zwanziger for the very useful 
discussions and communications. In addition, MP is greatful to the 
organizers and participants of the Cambridge workshop on gravity and Lorentz violation, 
that catalyzed the revision of this paper.
This work was supported in part by NSERC, Canada, and research at the Perimeter Institute
is supported in part by the Government of Canada 
through NSERC and by the Province of Ontario through MEDT.

\appendix

\section{Propagators and gauge fixing in Ho\v rava-Lifshitz gravity}
\label{app:propagators}
The action $\mathcal L_{\mathbf{1}}$ 
is identical in both Einstein's and Ho\v rava's theories if 
operators with Lifshitz-dimension higher than $6$ are ignored.
Both $\mathcal L_{\mathbf{1}}$ and $\mathcal L_{\mathbf{0}}$ contain
gauge symmetries generated by \eqref{eq:vector_gauge} and \eqref{eq:scalar_gauge},
and one must choose a gauge fixing scheme to
derive the propagators for vector and scalar gravitons.

The simplest gauge condition\footnote{As a matter of additional check, we 
have also performed calculations in the generalized 
$R_\xi$ gauge for the spin-1 gravitons, when 
${\cal L} = -\fr{1}{2\xi} (\dot n_i - \alpha \Delta V_i)^2$ 
term is added to the action. Explicit 
calculations of $c_{\rm scalar}$ and $c_{\rm photon}$ can be carried out, 
and the result for their difference shows complete 
independence on the choice of $\xi$ and $\alpha$ parameters.}
 to choose that respects the Lifshitz symmetry
at large momentum would be $n_i=0$. In this gauge, we easily derive the
propagators
\begin{equation}
\vev{V_i^\T\, V_j^\T}=-\frac{\delta_{ij}-k_i k_j/\vec k^2}{\omega^2\vec k^2}
\,,\qquad
\vev{n_i^\T\, \ast}=0\,,
\end{equation}
for vector-gravitons, and, at large $\vec k$, 
\begin{equation}
\label{eq:spin0_prop}
\begin{split}
\vev{\sigma\,\sigma}=&
-\frac{\tilde\lambda}
{\omega^2-\alpha\tilde\lambda\LambdaHL^{-4}\vec k^6}\,,\quad
\vev{\sigma\,\tau}=
-\frac{\tilde\lambda-1}
{\omega^2-\alpha\tilde\lambda\LambdaHL^{-4}\vec k^6}\,,\quad
\vev{\sigma\, n}=
\frac{c}{b}\, \frac{\tilde\lambda}
{\omega^2-\alpha\tilde\lambda\LambdaHL^{-4}\vec k^6}\,,\\
\vev{\varphi\, \ast}=&0\,,
\end{split}
\end{equation}
for spin-0 gravitons. We have defined the parameters
\begin{equation}
\alpha\equiv a-\frac{c^2}{b}\,,\qquad 
\tilde\lambda\equiv\frac{\lambda-1}{3\lambda-1}
\end{equation}
in the above expressions, and omitted the correlations functions that
are irrelevant to our results.  In this gauge, some of the propagators are
singular and more involved regularization scheme is proposed
\cite{Baulieu:1998kx,Baulieu:2007zzb}, but those subtleties do not
complicate our calculation here since our final answer is manifestly
gauge-choice independent and any  divergences that may arise
in the individual loop diagram will be cancelled out in physical
quantities.  Or, if any doubts remain, one can also carry out
the calculation in the $R_\xi$ gauge where a gauge fixing
term $\mathcal L=-\frac{1}{2\xi}(\dot n_i-\alpha\Delta V_i)^2$
is introduced. All the propagators in that gauge will be
``healthy'' and the final answer not only is independent of
the choice of $\xi$ and $\alpha$ but also agrees with that found
in the $n_i=0$ gauge.  Explicit calculation shows that
\begin{equation}
\vev{v^\T_i\, v^{\T i}}=-\frac{2}{\vec k^2}\,,\quad
\vev{\dot\sigma\,\chi}=\frac{\lambda}{3\lambda-1}
\frac{\omega^2}{\omega^2-\alpha\tilde\lambda\LambdaHL^{-4}\vec k^6}.
\end{equation}

We would like to make a few more comments on these results.
We find a new scale emerging in the
spin-0 sector. The propagators do
exhibit anisotropic scaling properties with 
the Lifshitz critical exponent $z=3$, but with a new
Lifshitz scale related $\LambdaHL$ as
\begin{equation}
\label{eq:lambdahlprime}
\LambdaHL'=\left(\alpha\tilde\lambda\right)^{-\frac{1}{4}}\LambdaHL\,.
\end{equation}
Depending on the value of $\alpha$ and $\tilde\lambda$, it can be much 
higher, lower than or equal to $\LambdaHL$. We will refer to this 
scale as the induced Lifshitz scale for the spin-0 sector.

Of course the choice of parameters $a,$ $b$ and $c$,
including their signs, may have direct consequences for the stability and
strong coupling problems in the gravity sector.
It is well known by analyzing the case for pure gravity
that $\tilde\lambda>0$ is a necessary condition to avoid 
ghosts. This cannot be immediately seen from the above 
propagators, as we have not fully diagonalized the action. 
We would not dwell on this issue further and simply
assume that  there exists reasonable choices of parameters, 
so that the theory is well defined.

To remove all quadratic divergence in the loop-induced
Lorentz violation effect observed in Sec. \ref{sec:LV},
we introduce the additional term in the theory:
\begin{equation}
\mathcal L'=\frac{2}{\Lambda^2}\left(\nabla_i K^{ij}\right)
\left(\nabla^k K_{kj}\right)
=\frac{1}{2\Lambda^2}v^\T_i \Delta^2 v^{\T i}
-\frac{2}{\Lambda^2}\chi\Delta\chi\,.
\end{equation}
It is easily checked that the propagators become
\begin{equation}
\vev{V_i^\T\, V_j^\T}=-\frac{\delta_{ij}-k_i k_j/\vec k^2}{\omega^2
(\vec k^2+\Lambda^{-2}\vec k^4)}
\,,\qquad
\vev{n_i^\T\, \ast}=0\,,
\end{equation}
for vector-gravitons, and, at large $\vec k$, 
\begin{equation}
\label{eq:spin0_prop}
\begin{split}
\vev{\sigma\,\sigma}=&\frac{1}{(2\lambda-1)}\frac{1}
{\omega^2-\alpha(1-2\lambda)^{-1}\LambdaHL^{-4}\vec k^6}\,,\quad
\vev{\sigma\,\chi}=\frac{i\lambda}{(4\lambda-2)}
\frac{\Lambda^2\omega}
{\vec k^2(\omega^2-\alpha(1-2\lambda)^{-1}\LambdaHL^{-4}\vec k^6)}\,,\\
\vev{\sigma\, n}=&-\frac{c}{b}\, \frac{1}{(2\lambda-1)}\frac{1}
{\omega^2-\alpha(1-2\lambda)^{-1}\LambdaHL^{-4}\vec k^6}\,,
\end{split}
\end{equation}
for spin-0 gravitons. Due to the additional $\vec k^2$ suppression 
in the $\sigma$-$\chi$ correlator, only the very last term
in Eq. \eqref{eq:final} contributes to the logarithmic divergence
produed by the scalar-graviton loops.

\section{Additional details about Lifshitz type loop integrals}
\label{app:lifshitz_fermion}
When the internal propagators all share the same Liftshitz-type behavior with
the same exponent $z$, the loop integral can be easily cast into a normal
Feynman integral so that standard textbook formulae are directly applicable.
Take, as an example, the fermion 1-loop correction discussed in
section \ref{sec:fermion}:
\begin{equation}
K=-\frac{1}{4i (2\pi)^4 M^2}\int \ud ^4 k \;
\frac{F^{\mu\nu} F^{\alpha\beta}
	\tr\,\sigma_{\mu\nu}\dslash{\tilde k}\sigma_{\alpha\beta}
	\dslash{\tilde k}}{\tilde k^4}\,,
\end{equation}
where
\begin{equation}
(\tilde k^0,\, \tilde{\vec k})\equiv
(k^0, \,|k|^{z-1}\vec k/\LambdaHL^{z-1})\,.
\end{equation}
It is easy to verify that 
\begin{equation}
\ud^4 \tilde k=\frac{z |\vec k |^{3(z-1)}}{\LambdaHL^{3(z-1)}}\ud^4 k
=\frac{z |\tilde {\vec k}|^{3(1-1/z)}}{\LambdaHL^{3(1-1/z)}}\ud^4 k\,.
\end{equation}
Changing the loop integral variable from $\ud^4 k$ to $\ud ^4 \tilde k$, 
and droping the tilde for brevity, we have 
\begin{equation}
\label{eq:zeroth}
K=-\frac{\LambdaHL^{3(1-1/z)}}{4i (2\pi)^4 M^2}\int \ud ^4 k \quad
\frac{F^{\mu\nu} F^{\alpha\beta}
	\tr\,\sigma_{\mu\nu}\dslash{k}\sigma_{\alpha\beta}
	\dslash{k}}{z |\vec k|^{3(1-1/z)} k^4}\,.
\end{equation}
The following identity is also easily checked
\begin{equation}
\tr\,\sigma_{\mu\nu}\dslash{k}\sigma_{\alpha\beta}\dslash{k}
=4k^2 (g_{\mu\beta}g_{\nu\alpha}-g_{\mu\alpha}g_{\nu\beta})
+8(g_{\mu\alpha}k_\nu k_\beta-g_{\mu\beta}k_\nu k_\alpha+g_{\nu\beta}k_\mu k_\alpha
	-g_{\nu\alpha}k_\mu k_\beta).
\end{equation}
Full Lorentz symmetry emerges as $z=1$, in which case
each pair of $k_\alpha k_\beta$ in the above expressions can be replaced 
by $\frac{1}{4} k^2 g_{\alpha\beta}$ and consequently $K$ vanishes 
identically by simple symmetry considerations.

While $z>1$ and the Lorentz symmetry is broken, using parity and spatial 
rotational symmetry, one can still replace each pair of 
$k_\alpha k_\beta$ in the integral by $g_{00} f^t+\sum_i g_{ii} f^x$.
Here
\begin{equation}
f^t\equiv -\frac{8}{i(2\pi)^4}\int \ud^4k\, \frac{k_0^2}{z |\vec k|^{3(1-1/z)} k^4}\,,
\end{equation}
and
\begin{equation}
f^x\equiv \frac{8}{3i (2\pi)^4} \int \ud^4k\, \frac{|\vec k|^2}{z |\vec k|^{3(1-1/z)} k^4}\,.
\end{equation}
Therefore, 
\begin{equation}
\begin{split}
K=&-\frac{\LambdaHL^{3(1-1/z)}}{M^2}
\left(-\frac{f^t+3 f^x}{4} F_{\mu\nu} F^{\mu\nu}
+ f^t F_{\mu 0} F^{\mu 0} +\sum_i f^x F_{\mu i} F^{\mu i}\right)\\
	=&\frac{\LambdaHL^{3(1-1/z)}(f^t-f^x)}{2 M^2} 
	(\mathbf E^2+\mathbf B^2)\,.
\end{split}
\end{equation}

Most generally in the Euclidean signature,
we have the integral \cite{Anselmi:2007ri}
\begin{equation}
\begin{split}
I^A_{r,s}\equiv &\int \frac{\ud^{\hat D} \hat p}{(2\pi)^{\hat D}}
\int \frac{\ud^{\bar D}\bar p}{(2\pi)^{\bar D}}
\frac{(\hat p^2)^r \, (\bar p^2)^s}
{(\hat p^2+\bar p^2+m^2)^A}\\
=&m^{2(r+s)+\hat D+\bar D-2A}\;
\frac{\Gamma\left(s+\frac{\bar D}{2}\right)
	\Gamma\left(r+\frac{\hat D}{2}\right)
		\Gamma\left(A-r-s-\frac{\hat D+\bar D}{2}\right)}
{(4\pi)^{(\hat D+\bar D)/2}\Gamma(\hat D/2)\Gamma(\bar D/2)\Gamma(A)}\,.
\end{split}
\end{equation}
Choosing $\hat D=1$, $\bar D=3$, and a Wick rotation leads to
$f^t=\frac{8}{z}I^2_{1, 3/(2z)-3/2}$, and 
$f^x=\frac{8}{3z}I^2_{0, 3/(2z)-1/2}$. For both $f^t$ and $f^x$,
\begin{displaymath}
r+s=\frac{3}{2z}-\frac{1}{2}.
\end{displaymath}
Therefore the ratio of the two is immediately given by
\begin{equation}
\frac{f^t}{f^x}= \frac{3 \Gamma\left(\frac{3}{2}\right)
	\Gamma\left(\frac{3}{2z}\right)}
{\Gamma\left(\frac{1}{2}\right)
	\Gamma\left(\frac{3}{2z}+1\right)}\,.
\end{equation}
As $z=1$, $f^t=f^x$ as expected. $z=3$ is a particular 
interesting case where we find
\begin{equation}
f^t=3 f^x=\frac{1}{2\pi^2}\Gamma(0).
\end{equation}
$\Gamma(0)$ encodes the UV divergence in this
formula, which gives rise to a logarithmically divergent terms 
for both $f^t$ and $f^x$ if we use dimensional regularizations. 
The fact that they are different by a factor
of $3$ implies violation of the Lorentz symmetry.

\section{Two toy models of Lifshitz scalar-QED}
\label{app:toys}
To achieve  better understanding of the physics in Lifshitz-type gauge theories, 
we intend to work out
two different toy models, in which an ordinary complex scalar is coupled to a 
Lifshitz-type photon.  In order to retain analogy to the graviton radiative corrections 
to the kinetic terms of non-Lifshitz matter fields, we evaluate 
the mass renormalization of the complex scalar 
generated by the photon loops.  These loop integrals are also quadratically divergent
in ordinary QED and expected to become better convergent 
if the photon is Lifshtiz-like.

There are two way of ``Lifshitzising'' the photon. 
One can do so by breaking all the gauge symmetries as in
the following theory:
\begin{equation}
\mathcal L=-(\partial_\mu\phi-iA_\mu\phi)(\partial^\mu\phi+i A^\mu \phi^\dag)
	+\frac{1}{2}A^\mu\left\{
	\left[\square-(-\Delta)^z\LambdaHL^{-2(z-1)}\right]
	g_{\mu\nu}
	-\partial_\mu\partial_\nu\right\} A^\nu\,,
\end{equation}
where $z\ge 2$. We used the combination $(-\Delta)=\vec k^2$
since it is a positive-definite operator.
This theory appears like the standard scalar-QED if $\LambdaHL\rightarrow\infty$
but breaks gauge symmetry explicitly as long as $\LambdaHL$ is finite. 
In this theory there is no need of gauge fixing and 
the propagators of $A_{\mu}$ is given by:
\begin{equation}
\vev{A_\mu A_\nu}=-\frac{g_{\mu\nu}+\frac{k_\mu k_\nu}
	{\LambdaHL^{-2(z-1)}\vec k^{2z}}}
{\omega^2-\vec k^2-\LambdaHL^{-2(z-1)}\vec k^{2z}}\,.
\end{equation}
There are two
relevant diagrams to evaluate for the mass renormalization of
$\phi$. The single-vertex diagram
corresponds to the following loop integral:
\begin{equation}
I_1=\frac{1}{(2\pi)^4}\int\frac{\ud \omega\ud^3 \vec k \,
	\left[4+\LambdaHL^{2(z-1)}(-\omega^2+\vec k^2)/
		\vec k^{2z}\right]}
{\omega^2-\vec k^2-\LambdaHL^{-2(z-1)}\vec k^{2z}}
\approx
\frac{3}{(2\pi)^4} \int\frac{\ud \omega\ud^3 \vec k}
{\omega^2-\LambdaHL^{-2(z-1)}\vec k^{2z}}\,.
\end{equation}
Here we have used the residue theorem and assumed that the dominant
part of the integral is contributed by the pole
at $\omega=\pm\,|\vec k|^z/\LambdaHL^{z-1}$. This integral is
logarithmically divergent if $z\ge 3$.

The double-vertex diagram consists of one scalar propagator and one photon propagator,
and in the limit of zero external momentum  is given by
the integral
\begin{equation}
\begin{aligned}
I_2=&\frac{1}{(2\pi)^4}\int\frac{\ud\omega\ud^3\vec k\,
k^2\left[1+k^2\LambdaHL^{2(z-1)}/\vec k^{2z}\right]} 
{(\omega^2-\vec k^2)
	\left[\omega^2-\vec k^2-\lambda_L^{-2(z-1)}\vec k^{2z}\right]}\,.
\end{aligned}
\end{equation}
This integral is finite as long as $z\ge 2$.

Therefore, in this toy model
the mass renormalization of $\phi$ is only linearly divergent
if $z=2$, logarithmically if $z=3$, and finite if $z>3$. 

We will now examine a different toy model which is much closer in spirit 
to Ho\v rava's theory of gravity. We would Lifshitzise
photon without breaking the gauge symmetry.  Consider the Lagrangian
\begin{equation}
\mathcal L=-\frac{1}{2} F_{0i} F^{0i}
-\frac{1}{4\LambdaHL^{2(z-1)}} F_{ij}(-\Delta)^{z-1} F^{ij}\,.
\end{equation}
Similar to ADM formalism in Lifshitz gravity, we separate
the variables $A_0$ and $A_i\equiv A_i^\T+\partial_i\varphi$ and
rewrite the action as
\begin{equation}
\mathcal L=
-\frac{1}{2}A^\T_i\left[\partial_t^2+\LambdaHL^{-2(z-1)}(-\Delta)^z\right]
A^{\T i}
-\frac{1}{2}(A^0+\dot\varphi)\Delta(A^0+\dot\varphi)\,.
\end{equation}
This expression makes explicit the gauge symmetry
\begin{equation}
A^0\rightarrow A^0-\dot\omega\,,\qquad 
\varphi\rightarrow\varphi+\omega\,,
\end{equation}
which is nothing but the original gauge symmetry 
$A_\mu\rightarrow A_\mu+\partial_\mu\omega$.
We would like to compute the mass renormalization for
the complex scalar in this model as well. Clearly
\begin{equation}
I_2=\frac{1}{(2\pi)^4}
\int\ud^4 k \; k^\mu k^\nu \vev{A_\mu\, A_\nu}\vev{\phi\,\phi^\dag}\,,
\end{equation}
and
\begin{equation}
I_1=-\frac{1}{(2\pi)^4}\int\ud^4 k\; g^{\mu\nu}\vev{A_\mu\, A_\nu}\,.
\end{equation}
Therefore, the sum
\begin{equation}
I_1+I_2=-\frac{1}{(2\pi)^4}
\int\ud^4 k\; \left(g^{\mu\nu}-\frac{k^\mu k^\nu}{k^2}\right)
\vev{A_\mu\, A_\nu}
\end{equation}
picks up only the gauge independent part of the 
photon propagator automatically.
We can choose any gauge that we like to evaluate 
these integrals.  For example, in the $A^0=0$ gauge, analogous to
the $n_i=0$ gauge in gravity, the photon propagators are
\begin{equation}
\begin{aligned}
\vev{A_i\, A_j}
=-\frac{1}{\omega^2-\LambdaHL^{-2(z-1)}\vec k^{2z}}\left(
\delta_{ij}-\frac{k_i k_j}{\vec k^2}\right)
-\frac{k_i k_j}{\omega^2\vec k^2}\,,\quad
\vev{A^0\,A^0}=\vev{A^0\,A_i}=0\,.
\end{aligned}
\end{equation}
Therefore,
\begin{equation}
\label{eq:mass_qed}
I_1+I_2=
\frac{2}{(2\pi)^4}
\int\frac{\ud\omega\ud^3\vec k}{\omega^2-\LambdaHL^{-2(z-1)}\vec k^{2z}}
+\frac{1}{(2\pi)^4}\int\frac{\ud\omega\ud^3\vec k}{\omega^2-\vec k^2}\,.
\end{equation}
Just as we have observed in the case of Ho\v rava type gravity,
this result, for $z=3$, contains both logarithmic and quadratic divergences.
The difference is that it is manifestly gauge independent in this simple toy model.
When $z=1$, 
$I_1+I_2=\frac{3}{(2\pi)^4}\int\ud\omega\ud^3 k\, (\omega^2-\vec k^2)^{-1}$, 
recovering the standard scalar-QED result.

\end{document}